# Selective Synthesis Combined with Chemical Separation of Single-Walled Carbon Nanotubes for Chirality Selection

Xiaolin Li, Xiaomin Tu, Sasa Zaric, Kevin Welsher, Won Seok Seo, Wei Zhao, Hongjie Dai*

*Department of Chemistry and Laboratory for Advanced Materials, Stanford University, Stanford, CA 94305, USA*
*Department of Chemistry, University of Arkansas, Little Rock, AR 72204, USA*

Single-walled carbon nanotubes (SWNTs) are potential materials for future nanoelectronics.[1,2] Since the electronic and optical properties of SWNTs strongly depend on tube diameter and chirality, obtaining SWNTs with narrow (n,m) chirality distribution by selective growth[3-5] or chemical separation[6-8] is important. Progresses have been made on selective growth of (6,5) SWNTs by chemical vapor deposition (CVD) of CO on CoMo catalyst (CoMoCAT)[3] and by ethanol CVD to a certain degree.[4] Methane CVD has been widely used for SWNT synthesis in bulk and on substrates for basic physics studies and device integrations,[9,10] but the nanotubes tend exhibit wide diameter distribution in the ~1-3nm range. SWNT growth in the d~1nm range with high diameter or chirality selectivity has been difficult by methane CVD thus far.

Here, we demonstrate that a new, bimetallic FeRu catalyst can afford SWNT growth with narrow diameter and chirality distribution in methane CVD. At 600°C, methane CVD on FeRu catalyst produced predominantly (6,5) SWNTs according to Uv-vis-NIR absorption and photoluminescence excitation/emission (PLE) spectroscopic characterization. At 850°C, the dominant semiconducting species produced are (8,4), (7,6) and (7,5) SWNTs, with much narrower diameter distributions than previous methane CVD grown SWNTs. Further, we show that narrow-diameter growth by methane CVD combined with chemical separation by ion exchange chromatography (IEC) facilitate achieving highly single-chirality enriched SWNT samples. This is demonstrated by obtaining SWNT samples highly enriched in the (8,4) tube and free of metallic species.

Methane CVD synthesis was preformed at various temperatures between 600°C and 850°C on a new silica-supported FeRu bimetallic catalyst (See supplementary information). We purified and suspended the synthesized SWNTs in surfactant solutions for characterizations. The UV-vis-NIR optical absorbance spectra (Fig.1) of our FeRu SWNTs grown at various temperatures showed one or several dominant peaks in the first van Hove $E_{11}$ range (900 to 1400nm) corresponding to SWNTs in the $d$~1nm region (see Supplementary Fig.S2 for atomic force microscopy images). Photoluminescence excitation/emission (PLE) spectroscopy was used to assign (m,n)[11] of SWNTs grown from FeRu catalyst. 600°C growth produced the narrowest (m,n) distribution with a dominant single peak corresponding to the (6,5) tube (Fig.2a). Higher growth temperature led to larger diameter tubes and wider (m,n) distribution of semiconducting tubes (Fig.2b,c). Growth at 700°C produced mainly (6,5) and (7,5) and (8,4) SWNTs (Fig.2b), while 850°C growth produced mainly (8,4), (7,5), (7,6) with few (6,5) tubes (Fig.2c). Metallic SWNTs existed the as-grown materials evident from the metallic $E_{11}$ absorption peaks in the 400-600nm region (Fig.1).

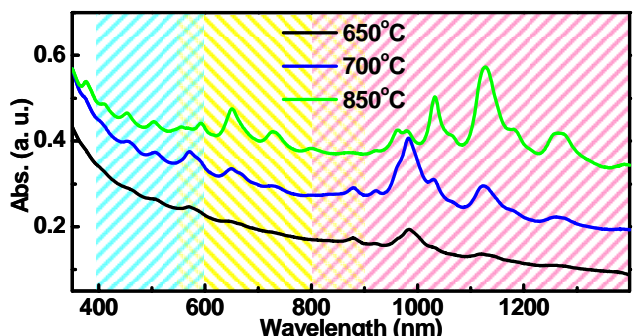

**Figure 1.** Uv-vis-NIR spectra of cholate suspended SWNTs grown by methane CVD on FeRu catalysts at various temperatures. The wavelength ranges of lowest energy semiconducting, second lowest semiconducting and lowest metallic absorption peaks are shaded pink, yellow and blue, respectively.

The FeRu SWNTs grown at 600°C are enriched in the (6,5) tube with high photoluminescence (PL) (Fig.2a) and weak signals from (m,n) = (7,6), (7,5) and (8,4) tubes (Fig.2a). It is difficult to analyze the absolute percentages of (6,5) tubes in the material due to the unknown optical absorption cross section and quantum yields for various (m,n) SWNTs[12,13,14] Note that recent experiment suggested no drastic chirality dependence for the PL brightness of SWNTs.[14] In a relative sense, we compared our material with CoMoCAT SWNTs known to be highly selective to the (6,5) tube.[3] We analyzed the (6,5)/(m,n) [for (m,n) = (7,6), (7,5) and (8,4) respectively] PL intensity ratios for both materials (Fig.3a and supplementary Fig.S1 for CoMoCAT), and found that the FeRu 600°C SWNTs are similarly enriched in the (6,5) SWNTs relative to other (m,n)'s compared to CoMoCAT.

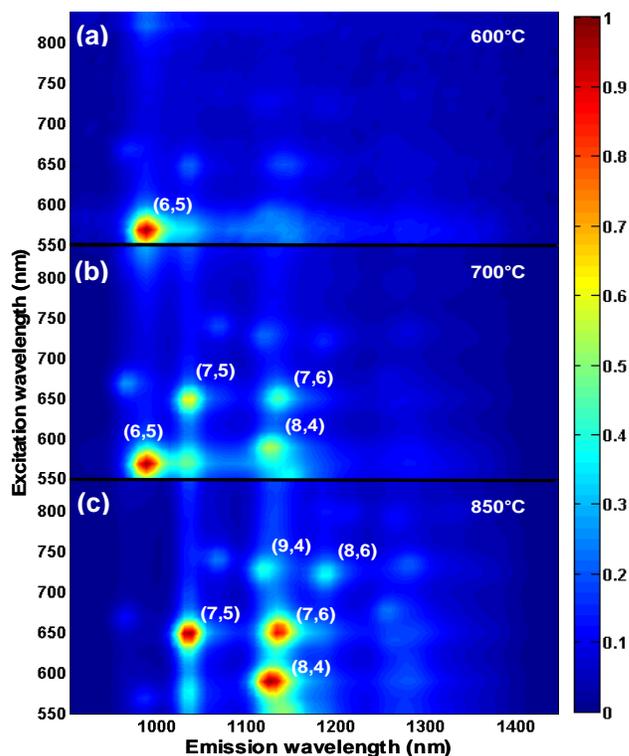

**Figure 2.** Contour plots of normalized photoluminescence emission intensities under various excitations for the Fe-Ru SWNTs grown at various temperatures. (m,n) is labeled at the right corner of each peak.

For FeRu SWNTs grown at 850°C, the (8,4) SWNT exhibited the highest PL intensity over other (m,n)'s (Fig.2c). We compared this material with Hipco SWNTs known to be broad in (m,n) distribution. We found much higher (8,4)/(6,5) and (8,4)/(m,n) [(m,n) = (7,6), (9,4)] PL intensity ratios for FeRu 850°C material than for Hipco (Fig.3b and supplementary Fig.S1 for Hipco). This suggested that compared to Hipco, the 850°C FeRu grown SWNTs are more selective to (8,4) SWNT than to (6,5) and (7,6) and (9,4) SWNTs. Note that we compared (8,4) tube to (7,6) and (9,4) tubes due to their similar diameters and $E_{11}$ emission peak positions (Fig.2c). Thus, the 850°C FeRu methane CVD produced abundant chiral (8,4) SWNT (chiral angle = 19.2°) away from arm-chair configuration. These growth results of narrow diameter distribution differ from typical previous methane CVD growth, suggesting that the FeRu catalyst, used here for the first time for CVD of nanotubes, might be unique and responsible for the growth results.

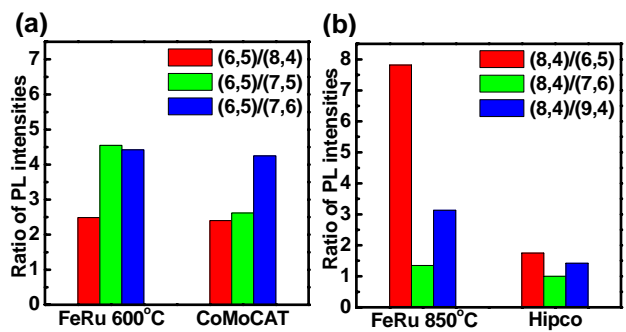

*Figure 3.* (a) PL intensity ratios between the (6,5) SWNT and the (8,4), (7,5) and (7,6) tubes in 600°C grown FeRu tubes vs. commercial CoMoCAT tubes. (b) PL intensity ratios between the (8,4) SWNT and (6,5), (7,6) and (9,4) tubes in 850°C grown FeRu tubes vs. Hipco tubes.

For the same methane CVD growth condition with other catalysts such as Fe/silica and FeMo/silica, SWNTs with 1-3nm diameters are produced with little selectivity in diameter and chirality.[10] The current Fe-Ru/silica catalyst produces SWNTs with diameter <~1.2nm (Fig.S2) and certain chirality selectivity. We suggest that the FeRu catalyst is unique owing to the low Fe-Ru eutectic temperature (~600-650°C).[15] The intimate alloying of FeRu and strong Fe-Ru interactions afford small catalytic nanoparticles stable against high-temperature sintering for producing small diameter SWNTs. Indeed, transmission electron microscopy revealed much smaller FeRu alloy particles (average size <2nm) smaller than particles in pure Fe or Fe/Mo catalysts after CVD growth. The large populations of FeRu nanoclusters in the 1-2nm range stable under the methane CVD conditions are responsible for the observed growth results. Diameter and chirality distributions generally widen and shift to larger size at higher temperatures found by various methods.

For the 850°C grown SWNTs by FeRu with (8,4) SWNT enrichment, we carried out DNA functionalization and then ion exchange chromatography (IEC) separation (see Supp. Info.) known to give diameter separated semiconducting SWNTs.[7] We obtained a fraction with the (8,4) SWNT as the dominant species (Fig.4b) with few metallic tubes (few peaks in 400-600nm range in Fig.4a). Compared to the as-grown material, the smaller diameter (6,5) and (7,5) tubes were eliminated by 3 and 5 fold respectively, and the similar diameter (7,6) and (9,4) tubes were reduced by 2 and 3 fold respectively. Thus, narrow diameter distribution growth and chemical separation combined facilitated obtaining highly enriched single-chirality SWNTs samples.

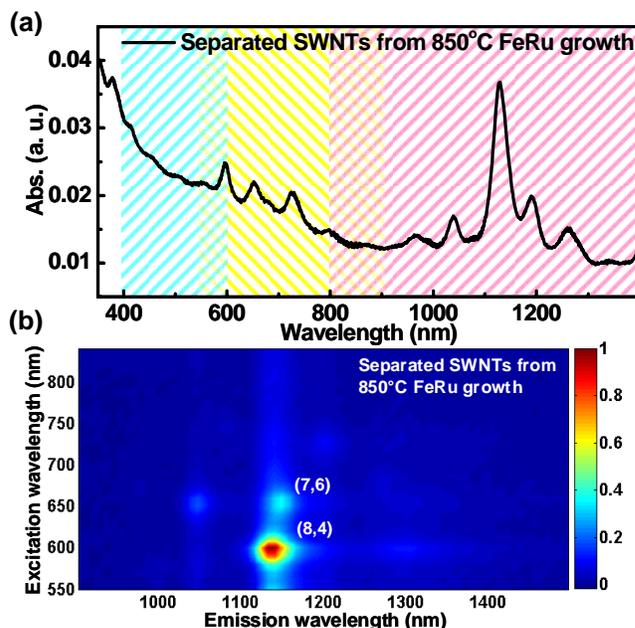

*Figure 4.* (a) UV-vis-IR and (b) PLE spectra of ion-exchange-chromatography IEC separated SWNTs from 850°C FeRu growth.

In summary, we developed a unique FeRu catalyst for selectively growth of SWNTs. This combined with chemical separation of the as-grown materials led to nearly single-chirality SWNT materials.

**Acknowledgment.** We thank MARCO MSD, Intel and Stanford GCEP for support.

**Supporting Information Available:** Experimental details are available free of charge via the Internet at http://pubs.acs.org.

ABSTRACT FOR WEB PUBLICATION.

Single-walled carbon nanotubes (SWNTs) are potential materials for future nanoelectronics. Since the electronic and optical properties of SWNTs strongly depend on tube diameter and chirality, obtaining SWNTs with narrow (n,m) chirality distribution by selective growth or chemical separation has been an active area of research. Here, we demonstrate that a new, bimetallic FeRu catalyst affords SWNT growth with narrow diameter and chirality distribution in methane CVD. At 600°C, methane CVD on FeRu catalyst produced predominantly (6,5) SWNTs according to Uv-vis-NIR absorption and photoluminescence excitation/emission (PLE) spectroscopic characterization. At 850°C, the dominant semiconducting species produced are (8,4), (7,6) and (7,5) SWNTs, with much narrower distributions in diameter and chirality than materials grown by other catalysts. Further, we show that narrow-diameter/chirality growth combined with chemical separation by ion exchange chromatography (IEC) greatly facilitate achieving single-(m,n) SWNT samples, as demonstrated by obtaining highly enriched (8,4) SWNTs with near elimination of metallic SWNTs existing in the as-grown material.

TOC Entry

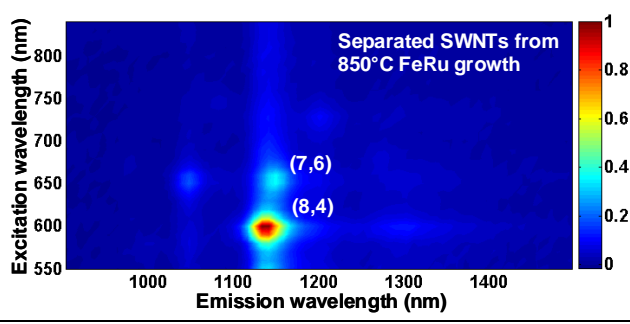